\documentclass[%
 reprint,
 amsmath,amssymb,
 aps,
prb,
]{revtex4-1}

\usepackage{graphicx}
\usepackage{dcolumn}
\usepackage{bm}

\newcommand{\ket}[1]{\left| #1 \right\rangle}
\newcommand{\bra}[1]{\left\langle #1 \right|}
\newcommand{\braket}[2]{\left\langle #1 | #2 \right\rangle}

\begin{document}

\preprint{APS/123-QED}

\title{Automated selection of the local potential for transferable pseudopotentials}

\author{Casey O. Barkan}
\author{Andrew M. Rappe}%
 \email{rappe@sas.upenn.edu}
\affiliation{%
 Department of Chemistry, University of Pennsylvania, Philadelphia, Pennsylvania 19104-6323, USA
}%

\date{\today}

\begin{abstract}
We develop an automated procedure to select the local potential of a separable pseudopotential that minimizes transferability errors for the isolated atom, and we show that this optimization leads to significant improvements in the accuracy of predicted solid-state properties. We present pseudopotentials for Y, In, and Sn. For these pseudopotentials, our method reduces solid-state errors by 88\% on average, as measured by the $\Delta$-factor test. These pseudopotentials are constructed in the Kleinman-Bylander form; however, our method is applicable to all separable pseudopotentials, such as ONCV pseudopotentials. We perform plane-wave convergence tests according to SSSP standards and show that the modifications to the local potential leave plane-wave convergence unchanged.
\end{abstract}

\pacs{Valid PACS appear here}
\maketitle


\section{Introduction}

The use of pseudopotentials in density functional theory (DFT) allows for a large reduction in computational cost with a small sacrifice in accuracy. In particular, pseudopotentials make the use of a plane-wave basis set feasible by replacing the rapid oscillations of valence wavefunctions in the core region with smooth pseudo-wavefunctions. The introduction of norm-conserving pseudopotentials (NCPPs) substantially improved transferability\cite{Hamann79p1494}, and many developments since then have led to large improvements in accuracy and efficiency\cite{Kleinman82p1425,Rappe90p1227,Troullier91p1993,Gonze91p8503,hamann2013optimized}. 
In addition, more complex methods have been developed to address some of the shortcomings of NCPPs, namely, ultrasoft pseudopotentials (USPPs) and the projector-augmented wave (PAW) method. However, the implementation of these methods can be quite complex\cite{hamann2013optimized}, and studies have shown that the accuracy of NCPPs can be comparable to that of USPPs and the PAW method\cite{holzwarth1997comparison,lejaeghere2016reproducibility}. As a result, NCPPs remain in widespread use.

The design of transferable NCPPs is essential for accurate DFT calculations. Traditionally, pseudopotential design has involved hand-picking parameters; however, optimization algorithms have recently been used to automate the selection of parameters for nonlocal pseudopotentials\cite{schlipf2015optimization,rivero2015systematic,hansel2015automated,sarkar2017evolutionary}. In this paper we revisit the designed-nonlocal (DNL) pseudopotential approach, in which the local potential of a separable norm-conserving pseudopotential is modified by adding an augmentation function in order to improve transferability \cite{Ramer99p12471}. We consider augmentation functions in the form of a sum of cosine functions, and use a conjugate gradient algorithm to select the coefficients in the sum that minimize transferability errors for the isolated atom. 

Some recent work on the use of optimization algorithms to design pseudopotentials has involved selecting parameters (such as core radii and kinetic energy truncation cutoff) that optimize the accuracy of solid-state properties\cite{schlipf2015optimization,rivero2015systematic,hansel2015automated,sarkar2017evolutionary}. Here, we take the different approach of optimizing transferability in the isolated atom only, and we show that this leads to dramatic improvements in the accuracy of solid-state properties. We measure pseudopotential accuracy in the solid-state with the $\Delta$-factor test\cite{lejaeghere2014error}, in which the equation of state of the elemental structure is compared between a pseudopotential calculation and an all-electron calculation. By optimizing only properties of the isolated atom, we ensure transferability by avoiding bias that can result from the particular choice of solid-state structures in the optimization.

In other previous studies, the local potential of separable pseudopotentials has been modified to improve agreement of the log-derivatives with all-electron results\cite{chou1992reformulation,garrity2014pseudopotentials}. While this approach ensures very good transferability for electron configurations very similar to the reference configuration, it does not guarantee that other configurations, such as oxidized or ionized states, will be reproduced accurately. In this paper, we modify the local potential to optimize pseudopotential accuracy over several configurations of the isolated atom in order to ensure greater transferability.

We demonstrate our automated DNL method by designing pseudopotentials for yttrium, indium, and tin, and test the accuracy of the designed pseudopotentials using the $\Delta$-factor test\cite{lejaeghere2014error}. 
Our method reduced the $\Delta$-factor by 98.6\%, 93.9\%, and 73.0\% for Y, In, and Sn, respectively. We note, however, that the $\Delta$-factor does not make reliable comparisons between pseudopotentials when the difference in $\Delta$ is small ($\lesssim 1$ meV/atom) \cite{lejaeghere2016reproducibility,prandini2018precision}. Furthermore, we demonstrate that the DNL approach can be used to tune a pseudopotential to reduce the $\Delta$-factor to nearly zero, but that the tuned pseudopotential has significantly higher transferability errors in the isolated atom than the pseudopotential designed using our DNL algorithm. Thus, a decrease in $\Delta$-factor does not necessarily translate to an improvement in transferability. Even so, the $\Delta$-factor represents the best benchmark that is currently available that enables systematic comparison of pseudopotentials across pseudopotential forms and implementations\cite{lejaeghere2014error,prandini2018precision}. 

Plane-wave convergence tests according to SSSP standards were also performed for the three pseudopotentials we present. All three pseudopotentials satisfy the SSSP precision convergence criteria at plane-wave cutoffs comparable to other NCPPs. We compare convergence of the pseudopotentials before and after optimization with the DNL approach, and find that the use of the DNL approach has no significant effect on plane-wave convergence.

The pseudopotentials presented in this paper are constructed in the single-projector Kleinman-Bylander (KB) form, though our approach can also be applied to multiple-projector pseudopotentials such as Optimized Norm-Conserving Vanderbilt (ONCV) pseudopotentials\cite{hamann2013optimized}. In its current implementation, our code can be used to optimize hybrid functional pseudopotentials using OPIUM, and it has already been shown that the DNL approach can improve transferability of hybrid functional pseudopotentials \cite{yang2018hybrid}. The pseudopotentials presented in this paper, however, use the PBE functional to provide a direct comparison with tabulated all-electron data\cite{lejaeghere2016reproducibility} used to evaluate the $\Delta$-factor. The code for our method is publicly available on the OPIUM pseudopotential generator website\cite{Opium}.

\section{\label{sec:method}Review of Designed-Nonlocal Approach}
The DNL approach involves modifying the local potential of a separable norm-conserving pseudopotential to improve transferability. The general form for a separable pseudopotential is
\begin{equation}
    \hat{V}^{\textrm{PS}}=V^{\textrm{loc}}(r) + \sum_l \hat{V}^{\textrm{NL}}_l
\end{equation}
where
\begin{equation}
    \hat{V}^{\textrm{NL}}_l = \sum_{i=1}^{N_{\textrm{proj}}} \ket{\chi_{li}}\frac{1}{b_{li}} \bra{\chi_{li}}
\end{equation}
and
\begin{equation}
   \ket{\chi_{li}}=(\epsilon_{li}-T-V^{\textrm{scr}}-V^{\textrm{loc}})\ket{\phi_{li}}.
\end{equation}
Here, $\phi_{li}/r$ and $\epsilon_{li}$ are the radial reference pseudo-wavefunctions and corresponding eigenvalues, $T=(-\frac{1}{2}\frac{d^2}{dr^2}+\frac{l(l+1)}{2r^2})$ is the kinetic energy operator, and $V^{\textrm{scr}}$ is the potential due to screening by the valence electrons in the pseudo-reference configuration. In this paper we use the Kleinman-Bylander form: a single projector per angular momentum ($N_{\textrm{proj}}=1$) and $b_{l1}=\braket{\phi_l}{\chi_l}$ \cite{Kleinman82p1425}. However the following method is applicable to all separable pseudopotentials, such as ONCV pseudopotentials. The reference pseudo-wavefunctions $\ket{\phi_{li}}$ were optimized for plane-wave convergence using the RRKJ method\cite{Rappe90p1227}.

To apply the DNL approach, we begin by setting $V^{\textrm{loc}}(r)$ equal to one of the semilocal potentials $V^{\textrm{SL}}_{\textrm{loc}}(r)$. For the pseudopotentials presented in this paper, the $l=0$ semilocal potential was used. Then, the local potential is modified with an augmentation function, $V^{\textrm{loc}}(r)=V^{\textrm{SL}}_{\textrm{loc}}(r)+A(r)$. Outside the core radius $r_c$, the local potential must be equal to the coulomb potential $Z_{\textrm{eff}}/r$, therefore $A(r)=0$ for $r>r_c$.

The effect of $A(r)$ can be seen by considering the action of the non-local projectors $\sum_l \hat{V_l}^{\textrm{NL}}$ on an arbitrary wavefunction $\ket{\psi}$. First, $\ket{\psi}$ can be decomposed into components according to angular momentum:
\begin{equation}
    \ket{\psi}=\sum_l \alpha_l \ket{\psi_l}.
\end{equation}
Then, the action of the non-local projectors on $\ket{\psi}$ is
\begin{equation}
    \sum_l \alpha_l \hat{V}^{\textrm{NL}}_l \ket{\psi_l} = \sum_{li} \beta_{li}[A(r)] \ket{\chi_{li}}
\end{equation}
where the coefficients $\beta_{li}[A(r)]$ are functionals of the augmentation function. In this sense, adjusting $A(r)$ allows one to design the effect of the non-local projectors. Of course, adjusting $A(r)$ has a direct impact on the local potential as well. Note that for $\ket{\psi}$ equal to one of the $\ket{\phi_{li}}$, the pseudopotential exactly reproduces the all-electron eigenvalues and wavefunctions for $r>r_c$, regardless of $A(r)$. The more dissimilar $\ket{\psi}$ is from the reference wavefunctions, the greater the effect of $A(r)$.

We define $A(r)$ as a series of cosine functions, with the constraint that $V^{loc}$  be continuous and differentiable at all $r$. In particular, we consider $A(r)$ of the form
\begin{equation}\label{eq:aug}
A(r) = 
\begin{cases}
a_0 + \sum_{n=1}^N a_n \cos{\frac{n \pi r}{L}} & r \leq L \\
0 & r > L
\end{cases}
\end{equation}
and we set $a_0$ according to
\begin{equation}\label{eq:constraint}
a_0 = \sum_{n=1}^N (-1)^{n-1} a_n
\end{equation}
so that $A(L)=0$ and $\frac{dA}{dr}|_{L}=0$. This ensures that $V^{\textrm{loc}}$ is continuous and differentiable at $r=L$. In this form, $A(r)$ is parametrized by the $N$ variables ${a_1,...,a_N}$. The outer bound $L$ is chosen to be slightly less than the smallest cutoff radius. For large $N$, $A(r)$ can approach any square-integrable function that satisfies the constraints at $r=L$. However, we have found that terms beyond $N=4$ provide only negligible improvements in transferability. The condition on $a_0$ ensuring continuity of $V^{\textrm{loc}}$ is not strictly necessary; however we have found some cases in which large discontinuities in $V^{\textrm{loc}}$ cause slower plane-wave convergence. For consistency, the pseudopotentials presented in this paper adhere to the constraint in Eq.~(\ref{eq:constraint}), though in general $a_0$ can be treated as an additional free parameter as long as thorough plane-wave convergence tests are performed.

\section{Method for DNL Optimization}

We begin by defining an objective function $F[A(r)]$ that maps $A(r)$ to a number that reflects the transferability errors of the pseudopotential. We then use a conjugate gradient algorithm\cite{shewchuk1994introduction} to find the coefficients $a_n$ in Eq.~(\ref{eq:aug}) that locally minimize the objective function.

Our objective function consists of comparisons between all-electron and pseudopotential results for several electron configurations of the isolated atom. These calculations are performed in the radial coordinate only, effectively assuming spherical symmetry. All isolated atom calculations were done using the OPIUM pseudopotential code\cite{Opium}. We define the objective function as
\begin{eqnarray}
F[A(r)] = \sum_{i=1}^C \bigg( \sum_{l}  a_{i,l} ({\Delta \epsilon_{i,l}})^2 + b_{i,l} ({\Delta t_{i,l}})^2  \nonumber \\
+ \sum_{j=i+1}^C c_{i,j}(\Delta E_{ij}^{\textrm{PS}} - \Delta E_{ij}^{\textrm{AE}})^2 \bigg)
\end{eqnarray}
where index $i$ indicates the $i$th electron configuration of the isolated atom, out of $C$ selected electron configurations, and the index $l$ indicates angular momentum channel. $\Delta \epsilon_{l,i}$ is the difference between the pseudo- and all-electron eigenvalue for the ($i$,$l$)th wavefunction. $\Delta t_{i,l}$ is the difference between the pseudo- and all-electron tail norm, defined as 
$t_{i,l}=(\int_{r_c}^{\infty} |\phi_{i,l}|^2 dr)^{1/2}$. $\Delta E^{\textrm{PS}}_{ij}$ and $\Delta E^{\textrm{AE}}_{ij}$ are the energy difference between configurations $i$ and $j$ as calculated by a pseudopotential calculation and all-electron calculation, respectively. The weights $a_{i,l}$, $b_{i,l}$, and $c_{i,j}$ can be chosen to give priority to certain properties or configurations. We set $a_{i,l}=1 \text{ (mRy)}^{-2}$, $b_{i,l}=10^6$, and $c_{i,j}=1 \text{ (mRy)}^{-2}$ for all $i$, $j$, and $l$. In the future, a statistical analysis could be performed to optimize these weights. In cases where semi-core is used, we include only the eigenvalues, tail norms, and energy differences corresponding to the valence wavefunction in the definition of $F[A(r)]$. Regarding the choice of electron configurations, we have had success using 3 or 4 excited states, and often an ionized state. The configurations used for the three pseudopotentials presented in this paper are shown in Table~\ref{tab:configs}.

\begin{table}
\caption{\label{tab:configs}Reference configurations used to construct the 3 pseudopotentials, and the test configurations used to define the objective function for each pseudopotential. Y was constructed with semi-core states. The $r_c$ values are the core cutoff radii for each angular momentum channel.}
\begin{ruledtabular}
\begin{tabular}{llllllll}
Atom & \multicolumn{2}{c}{Reference} & \multicolumn{5}{c}{Test configurations} \\
 & filling & $r_c$ (bohr) & 1 & 2 & 3 & 4 & 5 \\  \hline
Y&4s2&1.58 &4s2 & 4s2 & 4s2 & 4s2 & 4s2  \\
  &4p6&1.73 &4p6 & 4p6 & 4p6 & 4p6 & 4p6  \\
  &4d0&1.78 &4d1 & 4d2 & 4d0.9&4d0.5&4d0  \\
  &   & &5s2 & 5s1 & 5s1.9&5s1.4&5s2  \\
  &   & &5p0 & 5p0 & 5p0.2&5p0.1&5p0  \\ \hline
   
 In&4d10&2.09 &4d9.5 & 4d10 & 4d9.5 &  &  \\
   &5s2&2.18  &5s2   & 5s1  & 5s1.5   &  &  \\
   &5p1&2.06  &5p0.5 & 5p2  & 5p1 &  &  \\ \hline
   
 Sn&4d10&2.09 &4d10 & 4d9.5 & 4d9.5 &  &  \\
   &5s2&2.25  &5s1   & 5s2  & 5s1.5   &  &  \\
   &5p2&2.00  &5p3 & 5p2.5  & 5p3 &  &  \\ 
\end{tabular}
\end{ruledtabular}
\end{table}

We emphasize that this objective function uses comparisons of pseudo- and all-electron properties of isolated atoms only. This approach avoids bias that can arise from fitting to specific solid-state properties (although the DNL approach can be used to tune to specific solid-state properties, as we show in Sec.~\ref{sec:RD}). As an additional advantage, evaluation of the objective function is very rapid ($\approx1$ second on a single 3.3GHz core) compared to the time required to run solid-state calculations.

In conducting the optimization of $A(r)$, we begin with one term ($N=1$). We then add additional terms incrementally, always using the optimum from the ($N-1$)-term case as the starting point in the $N$-term optimization. In its current implementation, our method holds all pseudopotential input parameters fixed and adjusts only the augmentation function. In future work, this optimization method could be incorporated into a more comprehensive optimization strategy.

\begin{figure}
\includegraphics[width=5.5cm]{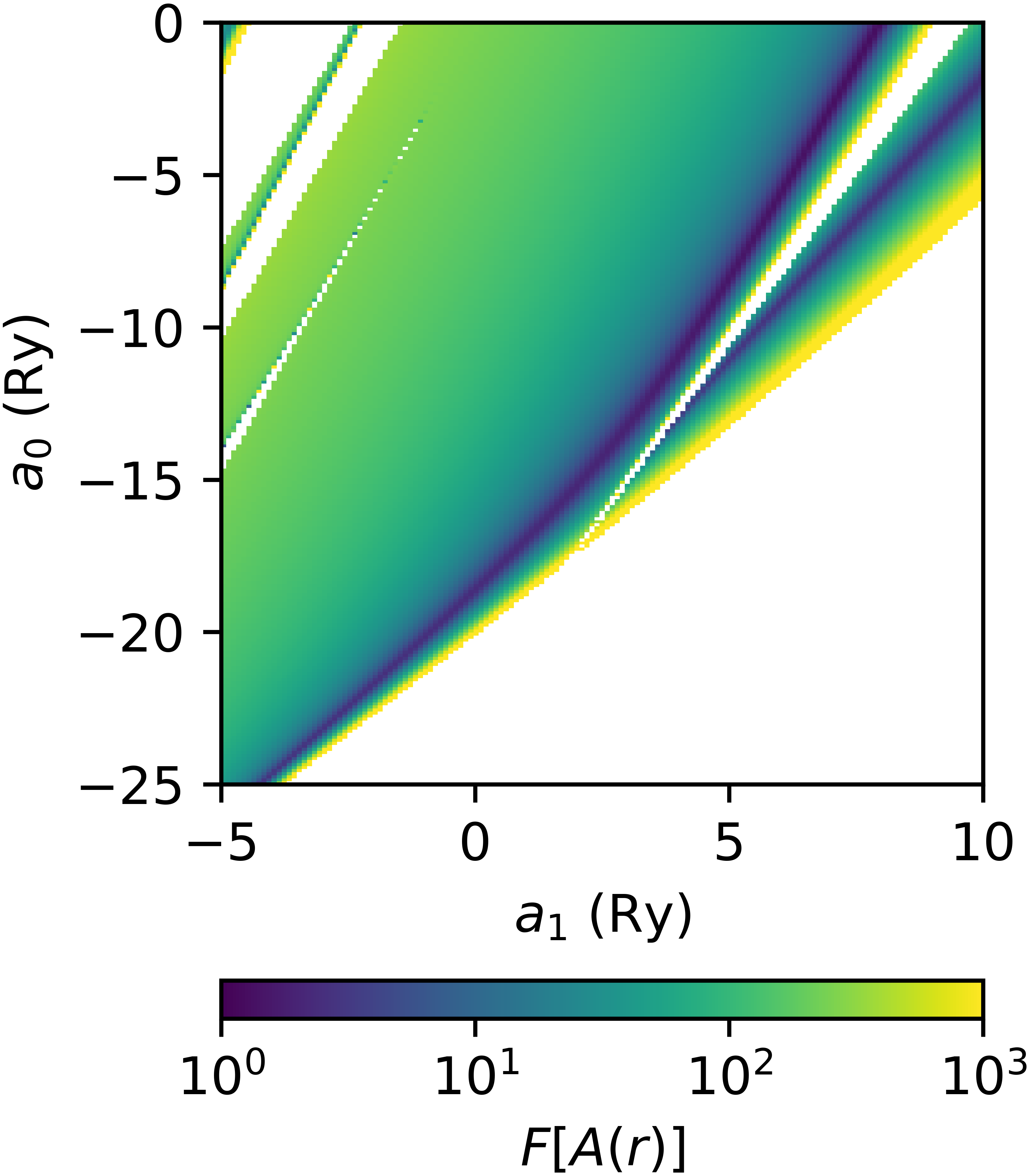}
\caption{\label{fig:2D}Plot of $F[A(r)]$ as a function of $a_0$ and $a_1$ (see Eq.~\ref{eq:aug}) for In. The presence of multiple local minima and deep valleys is a feature common to the objective function for all elements we've tested. White regions represent $A(r)$ which gave rise to a ghost state, or for which SCF convergence was not achieved during pseudopotential construction or during the calculation of test configurations.}
\end{figure}

It is instructive to plot the objective function as a function of $A(r)$ for the $N=2$ case. Such a plot for In, shown in Fig.~\ref{fig:2D}, illustrates two features that are common to all elements we have tried. First, it is clear that there are multiple local minima; in this paper, however, we do not attempt global minimization. Second, there are deep valleys in which the value of $F[A(r)]$ is roughly constant. This motivates the use of a conjugate gradient algorithm in order to avoid the so-called `zig-zag' problem\cite{shewchuk1994introduction}. The existence of these valleys also raises the concern that the algorithm will select extreme values for $a_n$, even while there exist less-extreme values for which $F[A(r)]$ is only negligibly increased. Although the use of only isolated atom properties in the objective function avoids bias, the objective function may neglect to account for the effect of $A(r)$ on neighboring atoms. This effect is typically small; however, for large $|a_n|$ it will not be negligible. In such a case, $F[A(r)]$ is no longer a complete measure of transferability. Thus, it is essential to test the pseudopotential accuracy in the solid-state in order to avoid this possibility.

To evaluate the quality of our pseudopotentials in solid-state calculations, we use the $\Delta$-factor test\cite{lejaeghere2014error}, in which the equation of state of the elemental structure is compared between pseudopotential and all-electron calculations. The $\Delta$-factor is defined as
\begin{equation}
    \Delta = \bigg( \frac{\int [\Delta E(V)]^2dV}{\Delta V} \bigg) ^{1/2}
\end{equation}
where $\Delta E(V)=E^{\textrm{AE}}(V)-E^{\textrm{PS}}(V)$, and where $E^{\textrm{AE}}(V)$ and $E^{\textrm{PS}}(V)$ are the energies per atom given by the third-order Birch-Murnaghan equation of state\cite{lejaeghere2016reproducibility} for the lowest energy structure of the given element, as calculated from an all-electron and pseudopotential calculation, respectively. Because only relative energies between states can be meaningfully compared between all-electron and pseudopotential calculations, $E^{\textrm{AE}}(V)$ and $E^{\textrm{PS}}(V)$ are set to equal $0$ at the equilibrium volume for the all-electron and pseudopotential calculations, respectively. The range of integration is from 94\% to 106\% of the all-electron equilibrium volume, and therefore $\Delta V=0.12 V^{\textrm{AE}}$. In order to calculate $\Delta$, we have used the Quantum Espresso 6.1 package for the pseudopotential calculations\cite{QE-2017,QE-2009} and we have used previously-published all-electron data\cite{lejaeghere2016reproducibility} calculated with the Wien2k package \cite{blaha2001wien2k}. To include the effects of relativity, we use scalar-relativistic pseudopotentials\cite{Bachelet82p2103}. To calculate the $\Delta$-factor we adopt the calculation parameters used in the SSSP verification tests\cite{prandini2018precision}: a k-point grid of $20\times20\times20$, a plane-wave cutoff of 200~Ry and Marzari-Vanderbilt smearing of 0.002 Ry.

Our results, discussed in Sec.~\ref{sec:RD}, support prior work demonstrating that pseudopotential accuracy in the isolated atom is a good predictor of accuracy in the solid-state\cite{Grinberg01p201102}.

\section{\label{sec:RD}Results and Discussion}

\begin{figure*}
\includegraphics[width=16.5cm]{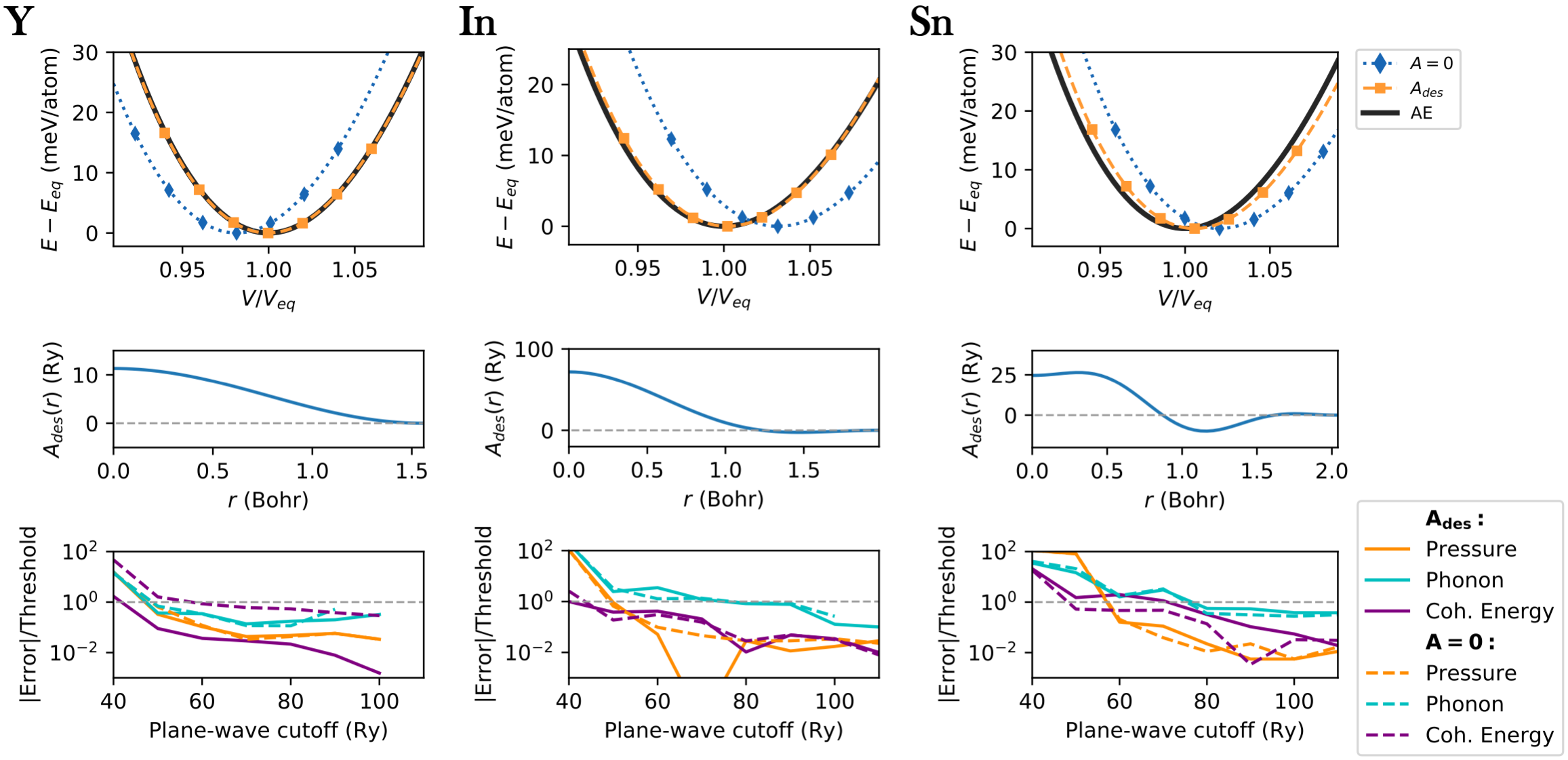}
\caption{\label{fig:EOS}Equation-of-state and plane-wave convergence test results for Y (left), In (center), and Sn (right). \textbf{Top panel:} Equation of state (EOS) curves for the undesigned ($A(r)=0$) and designed ($A_{\textrm{des}}$) pseudopotentials, compared to the all-electron EOS curve. \textbf{Middle panel:} graph of the designed augmentation function $A_{\textrm{des}}(r)$. \textbf{Bottom panel:} Results of the SSSP plane-wave convergence tests for pressure, phonons, and cohesive energy. Convergence errors are plotted as a fraction of the corresponding convergence threshold. Band structure convergence tests were also performed, and the designed and undesigned pseudopotentials for all three elements achieved the SSSP convergence criteria for band structure at 50 Ry plane-wave cutoff.}
\end{figure*}

We applied our automated-DNL method to three elements: Y, In, and Sn. By minimizing $F[A(r)]$, the $\Delta$-factor was reduced by 98.6\%, 93.9\%, and 73.0\% for Y, In, and Sn, respectively. Table~\ref{tab:resultsSummary} lists the values of $F[A(r)]$ and $\Delta$ for the undesigned ($A(r)=0$) and designed ($A(r)=A_{\textrm{des}}$) pseudopotentials for these elements. Fig.~\ref{fig:EOS} shows a comparison of the equation of state (EOS) curves between the Wien2k all-electron standard\cite{lejaeghere2016reproducibility,blaha2001wien2k}, the designed pseudopotential, and the undesigned pseudopotential. The optimal augmentation function $A_{\textrm{des}}$ for each element is also graphed. The results of plane-wave convergence tests are also shown, as is discussed further below.

\begin{table}
\caption{\label{tab:resultsSummary}Comparison of pseudopotentials with designed $A_{\textrm{des}}(r)$ to pseudopotentials with $A(r)=0$. Our method reduced the $\Delta$-factor by 98.6\%, 93.6\%, and 73.0\% for Y, In, and Sn, respectively.}
\begin{ruledtabular}
\begin{tabular}{ldddd}
 &\multicolumn{2}{c}{$A(r)=0$}&\multicolumn{2}{c}{$A(r)=A_{\textrm{des}}$}\\
 Atom &  \multicolumn{1}{c}{$F[A=0]$} &  \multicolumn{1}{c}{$\Delta$ (meV)} &  \multicolumn{1}{c}{$F[A_{\textrm{des}}]$} &  \multicolumn{1}{c}{$\Delta$ (meV)} \\ \hline
 Y&     1912.78&     5.61 &     1.21&         0.08 \\
 In&    44.41&     7.48 &     4.77&         0.46 \\
 Sn&    233.46&     6.30 &     4.96&         1.70 \\
\end{tabular}
\end{ruledtabular}
\end{table}

These results show that our approach can be used to reduce pseudopotential $\Delta$-factors to near or below 1 meV, below which the $\Delta$-factor test is not a reliable measure of accuracy\cite{lejaeghere2016reproducibility,prandini2018precision}. This demonstrates that single-projector pseudopotentials can have an accuracy comparable to multiple-projector pseudopotentials. Furthermore, this method can also be applied to multiple-projector pseudopotentials, as discussed above.

A key point illustrated in Fig.~\ref{fig:EOS} is that the improvements in $\Delta$ that result from optimizing $F[A(r)]$ are predominantly due to corrections in the lattice parameter. It is also noteworthy that, in some cases, $\Delta$ does not decrease monotonically with the number of terms $N+1$ in the augmentation function (despite $F[A(r)]$ necessarily decreasing with $N$). This was observed for the Sn pseudopotential we present, and we used the $\Delta$-factor to select between the pseudopotentials for each value of $N$. The number of terms in the augmentation function for the Y, In, and Sn pseudopotentials is 3, 4, and 5, respectively.

\begin{figure}
\includegraphics[width=5.0cm]{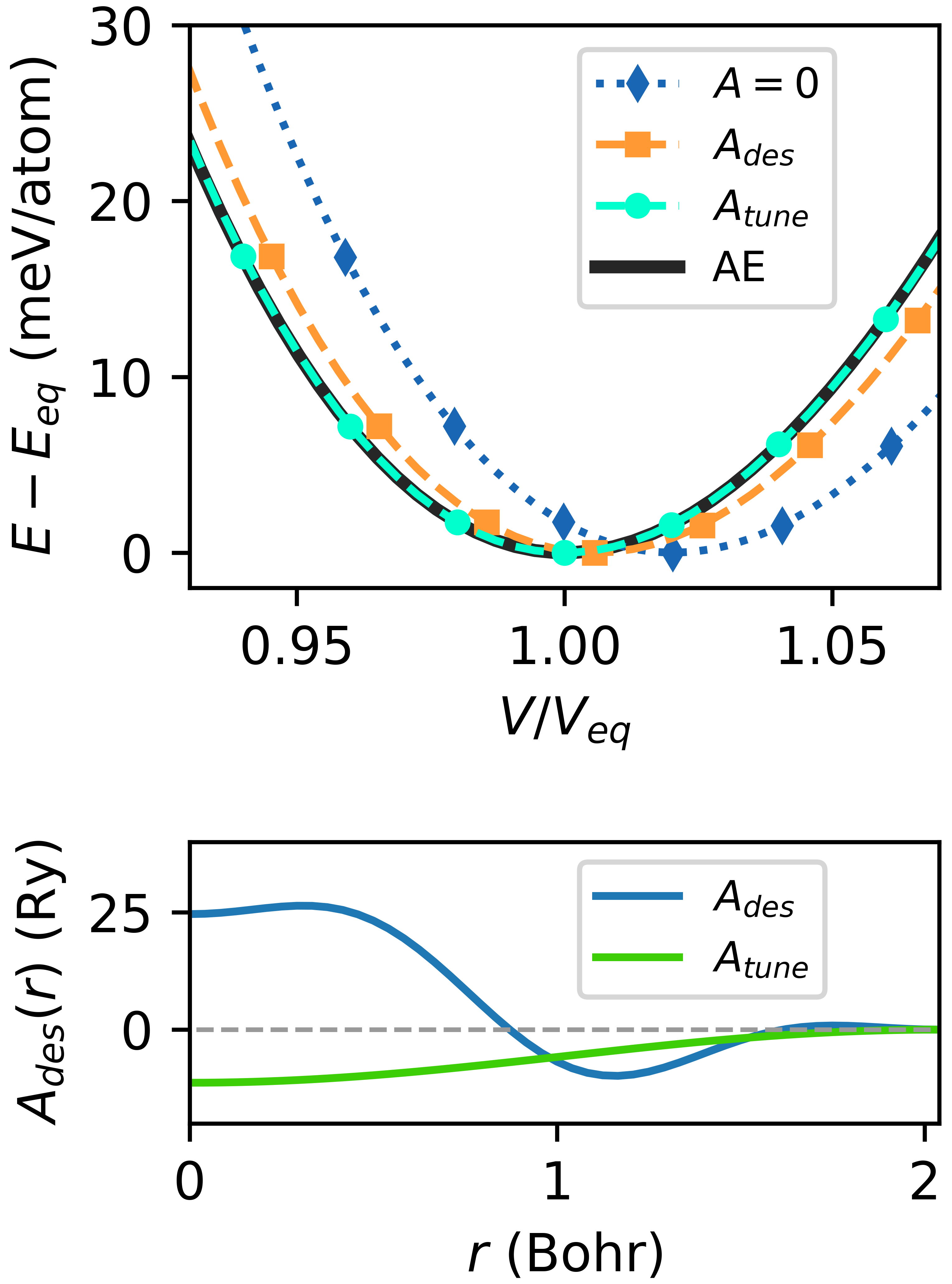}
\caption{\label{fig:tune}Equation of state curves for the tuned, designed, and undesigned pseudopotentials for Sn. The tuned pseudopotential achieves a much closer fit to the all-electron data. However, it performs worse in isolated-atom transferability tests.}
\end{figure}

Although we argue that including only isolated atom properties in the objective function avoids biasing towards particular solid-state properties, we show that the DNL approach can be used to tune a pseudopotential to reproduce the all-electron equation of state nearly perfectly. In particular, we tune a Sn pseudopotential to reproduce the all-electron lattice parameter by choosing the two-term $A(r)$ ($N=1$) that minimizes the difference between the pseudo- and all-electron lattice parameters. The $\Delta$-factor of this tuned pseudopotential is 0.04 meV, more than 40 times smaller than the $\Delta$-value of the designed Sn pseudopotential (1.70 meV) that we present in Table~\ref{tab:resultsSummary}. Fig.~\ref{fig:tune} shows the EOS curve for the tuned pseudopotential as well as the EOS curves for the pseudopotential with designed augmentation function $A_{\textrm{des}}$ and the pseudopotential with no augmentation function (undesigned). Although the tuned pseudopotential has much lower $\Delta$ than the designed pseudopotential, errors in the isolated-atom results, measured with the objective function $F[A(r)]$, are nearly three times greater for the tuned pseudopotential ($F[A_{\textrm{tune}}]=13.87$) than for the designed pseudopotential ($F[A_{\textrm{des}}]=5.3$). Thus, we do not recommend tuning as a method to generate transferable pseudopotentials. Rather, we conclude that the $\Delta$-factor, while a valuable tool, is not a comprehensive test of pseudopotential accuracy, and we instead focus on the minimization of errors in the isolated atom as our primary objective. 
In the future, more reliable testing may become possible through the development of more comprehensive benchmarks that retain the ability for systematic comparison across pseudopotential forms and implementations.

In addition to the $\Delta$-factor tests of pseudopotential accuracy, we performed thorough plane-wave convergence tests according to the SSSP standards\cite{prandini2018precision}. The three pseudopotentials we present require a plane-wave cutoff comparable to other norm-conserving pseudopotentials, and the use of the augmentation function was found to have a very small effect on convergence.

Plane-wave convergence was tested by examining the convergence of four solid-state properties specified in the SSSP tests: pressure, phonon frequencies, cohesive energy, and band structure. For each property, values at a selection of plane-wave cutoffs were compared to the converged values obtained using a 200 Ry plane-wave cutoff. Each test has a convergence error metric and corresponding convergence threshold. The plane-wave cutoff used in a DFT calculation should be greater than or equal to the cutoff at which all convergence error metrics are less than their corresponding threshold.

The metric for the pressure test is
\begin{equation}
    \delta V_{\textrm{press}} = \frac{V_{\textrm{BM}}(P[E_c]) - V_0}{V_0}
\end{equation}
where $V_0$ is the volume of the unit cell at the converged (200 Ry) cutoff, $P[E_c]$ is the pressure of the converged structure obtained from a calculation at a plane-wave cutoff of $E_c$. $V_{\textrm{BM}}(P)$ is the Birch-Murnaghan equation of state relating pressure to volume.

The metric for the phonon frequency test is
\begin{equation}
    \delta \bar{\omega} = \sqrt{ \frac{1}{N} \sum_i \Big( \frac{\omega_i[E_c] - \omega_i[200\textrm{ Ry}]}{\omega_i[200\textrm{ Ry}]}   \Big)^2 }
\end{equation}
where the sum is over all phonons of the structure, and $\omega_i[E_c]$ is the frequency of the $i$th phonon from a calculation at a cutoff of $E_c$. If the highest phonon frequency is less than 100cm$^{-1}$, an alternative metric $\delta \bar{\omega}'$ is used, defined as
\begin{equation}
    \delta \bar{\omega}' = \sqrt{ \frac{1}{N} \sum_i \big( \omega_i[E_c] - \omega_i[200\textrm{ Ry}]   \big)^2 }
\end{equation}

The cohesive energy metric $\delta E_{\textrm{coh}}$ is the deviation of the cohesive energy calculated at a plane-wave cutoff of $E_c$ from the converged cohesive energy.

The band structure metric is given by the $\eta_{10}$ value defined as
\begin{equation}
    \eta_{10} = \min_{\omega} \sqrt{ \frac{ \sum_{n\bm{k}} (\tilde{f}_{n\bm{k}})(\epsilon_{n\bm{k}}[E_c]-\epsilon_{n\bm{k}}[200\textrm{ Ry}] + \omega)^2 }{\sum_{n\bm{k}}\tilde{f}_{n\bm{k}} } }
\end{equation}
where
\begin{equation}
    \tilde{f}_{n\bm{k}} = \sqrt{f_{n\bm{k}}[E_c] f_{n\bm{k}}[200\textrm{ Ry}]}
\end{equation}
and where $f_{n\bm{k}}$ is the fillings of the $n$th band at k-point $\bm{k}$, given by the Fermi-Dirac distribution:
\begin{equation}
    f_{n\bm{k}}[E_c] = 1/(e^{(\epsilon_{n\bm{k}}[E_c]-\epsilon_F[E_c])/\sigma}+1)
\end{equation}
$\epsilon_F[E_c]$ is the Fermi energy of the system plus 10 eV, in order to include bands up to 10 eV above the Fermi energy in the convergence test. $\omega$ is a rigid shift in energy that accounts for any uniform offset in band energies between calculations at a plane-wave cutoff of $E_c$ and 200 Ry. $\sigma$ is the smearing, equal to 0.02 Ry. An additional convergence metric for band structure convergence is defined as $\max \eta_{10} = \max_{n\bm{k}}|\epsilon_{n\bm{k}}[E_c]-\epsilon_{n\bm{k}}[200\textrm{ Ry}] + \omega|$. A small $\max \eta_{10}$ ensures that the band structure converges well in all regions of the Brillouin zone.

The convergence thresholds specified by the SSSP precision criteria are $\delta V_{\textrm{press}} < 0.005$, $\delta \bar{\omega} < 0.01$ (or $\delta \bar{\omega}' < 1$~cm$^{-1}$ if all phonon frequencies are less than 100~cm$^{-1}$), $\delta E_{\textrm{coh}} < 2$~meV/atom, $\eta_{10} < 10$~meV, and $\max \eta_{10} < 20$~meV. The ratio of convergence metric to convergence threshold for the pressure, phonon, and cohesive energy tests are shown for a range of $E_c$ in Fig.~\ref{fig:EOS}. The results of the band structure convergence tests are not shown in Fig.~\ref{fig:EOS}, but the convergence threshold for both $\eta_{10}$ and $\max \eta_{10}$ are satisfied at $E_c=50$~Ry for Y, In, and Sn.

All three pseudopotentials we present converge at a plane-wave cutoff comparable to other norm-conserving pseudopotentials meeting the SSSP precision criteria: convergence is achieved at 50~Ry, 80~Ry, and 80~Ry for Y, In, and Sn, respectively. These strict convergence criteria are only necessary for calculations requiring very high accuracy. For less strict criteria, $E_c$ could be lowered substantially. Furthermore, the plane-wave optimization cutoffs used in the RRKJ optimization can be lowered to produce pseudopotentials that converge (according to less strict criteria) at much lower $E_c$, for use in calculations where the strict SSSP precision criteria are not necessary.

As is evident from Fig.~\ref{fig:EOS}, $A(r)$ has a small effect on plane-wave convergence. In some cases the designed pseudopotential has slightly better convergence, and in other cases slightly worse convergence, than the undesigned pseudopotential. For all three pseudopotentials we present, the plane-wave cutoff at which SSSP thresholds are satisfied is the same for the designed and undesigned pseudopotential.

\section{Conclusion}
We propose an automated method to select the local potential of a separable pseudopotential that maximizes transferability based on only isolated atom properties, and we show that this method leads to substantial improvements in the accuracy of solid-state properties, as measured with the $\Delta$-factor test. The $\Delta$-factor was reduced to near or below 1~meV for the Y, In, and Sn pseudopotentials that we present. We also show that the local potential can be tuned to reduce the $\Delta$-factor to nearly zero, but that doing so increases transferability errors of the isolated atom. Thorough plane-wave convergence tests were performed, and it was found that the use of the DNL approach has an insignificant impact on plane-wave convergence.

\begin{acknowledgments}
This work has been supported by the Department of Energy Office of Basic Energy Sciences, under grant number DE-FG02-07ER46431. The authors also
acknowledge computational support from the National
Energy Research Scientific Computing Center. COB acknowledges summer support from the Summer Undergraduate Research Group Grant program at the University of Pennsylvania.
\end{acknowledgments}

\bibliography{main}

\end{document}